\documentclass[aps,showkeys,groupedaddress,twocolumn,showpacs]{revtex4/revtex4}
\usepackage{graphicx}
\usepackage[ansinew]{inputenc}
\usepackage[tbtags]{amsmath}
\usepackage{amssymb}
\usepackage{epsfig}
\newcommand{\erfc}{\mbox{erfc}}
\newcommand{\erf}{\mbox{erf}}

\begin{document}
\title{When are Extreme Events the better predictable, the larger they are?}
\author{S. Hallerberg, H. Kantz}  
\affiliation{
Max Planck Institute for the Physics of Complex Systems\\
N\"othnitzer Str.\ 38, D 01187 Dresden, Germany\\
}

\date{\today}
\begin{abstract}
We investigate the predictability of extreme events in time series.  The focus of this work is to understand under which circumstances large events are better predictable than smaller events.
Therefore we use a simple prediction algorithm based on precursory structures which are identified using the maximum likelihood principle. 
Using the receiver operator characteristic curve as a measure for the quality
of predictions we find that the dependence on the event magnitude is closely linked
to the probability distribution function of the underlying stochastic
process. 
We evaluate this dependence on the probability distribution function analytically
and numerically. 
If we assume that the optimal precursory structures are used to make the
predictions, we find that large increments are better predictable if the
underlying stochastic process has a Gaussian probability distribution
function, whereas larger increments are harder to predict if the underlying
probability distribution function has a power law tail. 
In the case of an
exponential distribution function we find no significant dependence on the event magnitude.
Furthermore we compare these results with predictions of increments in
correlated data, namely, velocity increments of a free jet flow. The velocity
increments in the free jet flow are in dependence on the time scale
either asymptotically Gaussian or asymptotically exponential distributed. 
The numerical
results for predictions within free jet data are in good agreement with the previous analytical considerations for random numbers.  
\end{abstract}
\maketitle
\section{Introduction}
Systems with a complex time evolution, which generate
a great impact event from time to time, are ubiquitous. 
Examples include
fluctuations of prices for financial assets in economy with rare market
crashes, electrical activity of human brain with rare epileptic seizures,
seismic activity of the earth with rare earthquakes, changing weather
conditions with rare disastrous storms, and also fluctuations of online
diagnostics of technical machinery and networks with rare breakdowns or
blackouts. 
Due to the complexity of the systems mentioned, a complete
modeling is usually impossible, either due to the huge number of degrees of freedom involved, or due to a lack of precise knowledge about the governing
equations.
This is why one applies the framework of prediction via precursory structures
for such cases. The typical application for prediction with precursory
structures is a prediction of an event which occurs in the very near future,
i.e., on short timescales compared to the lifetime of the system.
A classical example for the search for precursory structures is the prediction
of earthquakes \cite{Jackson}. 
A more recently studied example is the short
term prediction of strong turbulent wind gusts, which can  destroy wind
turbines \cite{Physa, Euromech}. 

In a previous work \cite{Sarah2}, we studied the quality of predictions
analytically via precursory structures for increments in an AR(1) process and
numerically in a long-range correlated ARMA process. 
The long-range
correlations did not alter the general findings for Gaussian processes,
namely, that larger events are better predictable. 

Furthermore we found other works which report the same effect for earthquake
prediction \cite{Fatehme}, prediction of avalances in SOC-models
\cite{Sandpile} and in multiagent games \cite{Johnson1}.
In this contribution, we investigate the influence of the probability
distribution function (PDF) of the noise
term in detail by using not only Gaussian, but also exponential and power-law
distributed noise. 
This approach is also motivated by the book of Egans \cite{Egans}
which explains that receiver operator characteristics (ROC) obtained in signal detection problems can be
ordered families of functions in dependence on a parameter. 
We are now interested in learning how the behavior of these families of functions depends on the event magnitude and the distribution of the stochastic process,
if the ROC curve is used for evaluating the quality of predictions.\\
After defining the prediction scheme in Sec.\ \ref{pre} and the method for
measuring the quality of a prediction in Sec.\ \ref{roc},
 we explain in Sec.\ \ref{evsize} how to consider the influence on the event magnitude. 
In Sec.\ \ref{constr} we formulate a constraint, which has to be fulfilled in order to find a better predictability of larger (smaller) events. 
In the next section, we apply this constraint to compare the quality of
predictions of large increments within Gaussian (Sec.\ \ref{Gaussian}), exponential distributed
(Sec.\ \ref{symexpo}) and power-law distributed i.i.d.\ random numbers (Sec.\ \ref{powl}). 
We study the prediction of increments in free jet data in Sec.\ \ref{freejet}.
Conclusions appear in Sec.\ \ref{conclusions}.
\section{Definitions and setup\label{sec2}}
The considerations in this section are made for a time series \cite{Box-Jen,Brockwell}, i.e.\ , a set of measurements $x_n$ at discrete
times $t_n$, where $t_n = t_0 + n\Delta$ with a sampling
interval $\Delta$ and $ n \in \mathbb{N}$. 
The recording should contain sufficiently many extreme events so that we are able to extract statistical
information about them. 
We also assume that the event of
interest can be identified on the basis of the observations, e.g.\, by the
value of the observation function exceeding some threshold, by a sudden
increase, or by its variance exceeding some threshold.
 We express the presence (absence) of an event by using a binary variable $Y_{n+1}$.
\begin{eqnarray}
Y_{n+1} & = &\left\{  \begin{array}{ll}
1 \quad& \mbox{an event occurred at time} \\
& n+1 \\
 0 \quad& \mbox{no event occurred at time} \\
& n+1\end{array} \right. 
\end{eqnarray} 
\subsection{The choice of the precursor \label{pre}}
When we consider prediction via precursory structures ({\sl precursors},
or {\sl predictors}), we are typically in a situation where we assume that the
dynamics of the system under study has both, a deterministic and a stochastic
part. 
The deterministic part allows one to assume that there is a relation between
the event and its precursory structure which we can use for predictive
purposes. 
However, if the dynamic of the system was fully deterministic
there would be no need to predict via precursory structures, but we could
exploit our knowledge about the dynamical system as it is done, e.g., in weather forecasting. 

In this contribution we focus on the influence of the stochastic part of the
dynamics and assume therefore a very simple deterministic correlation between
event and precursor.%
The presence of this stochastic part determines that we cannot expect the precursor to preced {\em every} individual event. 
That is why we define a precursor in this context as a data structure which is {\em typically} preceding an event, allowing deviations from the given structure, but also allowing events without
preceeding structure. 

For reasons of simplicity the following considerations are made for precursors
in real space, i.e., structures in the time series. 
However, there is no reason not to apply the same ideas for precursory structures, which live in phase space.

In order to predict an event $Y_{n+1}$ occurring at the time $(n+1)$ we compare the last $k$ observations, to which we will refer as the {\sl precursory variable} 
\begin{equation} 
{\mathbf x}_{(n-k+1,n)}= (x_{n-k+1},x_{n-k+2}, ..., x_{n-1},x_n)
\end{equation}
with a specific precursory structure 
\begin{equation}
{\mathbf x}^{pre}= (x_{n-k+1}^{pre}, x_{n-k+2}^{pre}, ...,x_{n-1}^{pre}, x_n^{pre}).
\end{equation} 
Once the precursory structure  ${\mathbf x}_{pre}$ is determined, we give an
alarm for an event $Y_{n+1}=1$ when we find the precursory variable
${\mathbf x}_{(n-k+1,n)}$ inside the volume 
%
\begin{eqnarray}
V^{pre}(\delta,{\mathbf x}^{pre}) & = & \prod_{j=n-k+1}^{n} \left(x_j^{pre}-\frac{\delta}{2},x_j^{pre}+\frac{\delta}{2}\right). \label{vol}
\end{eqnarray}
%
There are different strategies to identify suitable precursory structures. 
We choose the precursor via maximizing a conditional probability which we refer to as the {\sl likelihood} \cite{likelihood}.
\footnote{In this contribution we use the name likelihood for the probability that an event follows a precursor ${\mathbf x}$. 
And the term {\sl aposterior pdf} for the probability to find a precursor ${\mathbf x}$ before of an already observed extreme event. 
Note that the names might be also used vice versa,  if one refers to the precursor as the previously observed information.}
The likelihood  
\begin{eqnarray}
L(Y_{n+1}=1|{\mathbf x}_{(n-k+1,n)}) = \frac{j(Y_{n+1}=1,{\mathbf
    x}_{(n-k+1,n)})}{\rho({\mathbf x}_{(n-k+1,n)})} \label{likely}
\end{eqnarray}
 provides the probability that an event $Y_{n+1}=1$ follows the precursor ${\mathbf x}_{(n-k+1,n)}$.
 It can be
calculated numerically by using the joint PDF $j((Y_{n+1}=1),{\mathbf
  x}_{(n-k+1,n)})$. 
Our prediction strategy consists of determining those values of each
component $x_i$ of ${\mathbf x}_{(n-k+1,n)}$ for which the likelihood is
maximal. 

This strategy to identify the optimal precursor represents a rather fundamental choice.
In more applied examples one looks for precursors which minimize or maximize
more sophisticated quantities, e.g., discriminant functions or loss matrices.
These quantities are usually functions of the posterior PDF or the
likelihood, but they take into account the additional demands of the
specific problem, e.g., minimizing the loss due to a false prediction.
The strategy studied in this contribution is thus fundamental in the
sense that it enters into many of the more sophisticated quantities which
were used for predictions and decision making.
\subsection{Testing for predictive power\label{roc}}
A common method to verify a hypothesis or to test the quality of a prediction
is the receiver operating characteristic curve (ROC) \cite{Swets1, Egans, Pepe}.
The idea of the ROC consists simply of comparing the rate of correctly
predicted events $r_{c}$ with the rate of false alarms $r_{f}$ by plotting
$r_c$ vs.\ $r_f$. 
The rate of correct predictions $r_c$ and the rate of false
alarms $r_f$ can be obtained by integrating the {\sl aposterior} PDFs $\rho({\mathbf
  x}_{(n-k+1,n)}|Y_{n+1}=1)$ and $\rho({\mathbf x}_{(n-k+1,n)}|Y_{n+1}=0)$ on the
precursory volume.
\begin{eqnarray}
r_c(\delta,{\mathbf x}^{pre}) & = & \int  \rho({\mathbf x}_{(n-k+1,n)}|Y_{n+1}=1)
dV^{pre}(\delta,{\mathbf x}^{pre}) \nonumber
\label{rcor}\\
\\
r_f(\delta,{\mathbf x}^{pre}) & = & \int  \rho({\mathbf x}_{(n-k+1,n)}|Y_{n+1}
=0) dV^{pre}(\delta,{\mathbf x}^{pre})\nonumber\\
\label{rf}
\end{eqnarray}
 Note that these rates are defined with respect to the total
numbers of events $Y_{n+1}=1$ and nonevents $Y_{n+1}=0$. 
Thus the relative frequency of events has no direct influence on the ROC, unlike on other measures of predictability, as e.g., the Brier score or the ignorance\cite{bandi}. 

Plotting $r_c$ vs $r_f$ for increasing values of $\delta$ one obtains a curve
in the unit-square of the $r_f$-$r_c$ plane (see, e.g., Fig.\ \ref{fig:rocgauss}).
The curve approaches the origin for $\delta \rightarrow0$ and the point $(1,1)$ in the limit $\delta \rightarrow \infty$, where $\delta$ accounts for the magnitude of the precursory volume $V_{pre}(\delta)$. 
The shape of the curve characterizes the significance of the prediction. A curve above the diagonal reveals that the corresponding strategy of prediction
is better than a random prediction which is characterized by the
diagonal. Furthermore we are interested in curves which converge as fast as
possible to $1$, since this scenario tells us that we reach the highest
possible rate of correct prediction without having a large rate of false
alarms.

That is why we use the so-called {\it likelihood ratio} as a summary index, to
quantify the ROC. 
For our inference problems the likelihood ratio
is identical to the slope $m$ of the ROC-curve at the vicinity of the origin
which implies $\delta \rightarrow 0 $. 
This region of the ROC is in particular interesting, since it corresponds to a low rate of false alarms.
The term likelihood ratio results from signal detection theory. 
In the context of signal detection theory, the term {\sl a posterior PDF} refers to the PDF, which we call likelihood in the context of predictions and vice versa.
This is due to the fact that the aim of signal detection is to identify a
signal which was already observed in the past, whereas predictions are made
about future events. 
Thus the \lq\lq likelihood ratio" is in our notation a ratio of a posterior PDFs.
\begin{eqnarray}
m & = & \frac{\Delta r_c}{\Delta r_f} \sim  \frac{\rho({\mathbf x}^{pre}|Y_{n+1}=1)}{\rho({\mathbf x}^{pre}|Y_{n+1}=0)} + \mathcal{O}(\delta)\label{defm}.
\end{eqnarray}
However, we will use the common name likelihood ratio throughout the text.
For other problems the name likelihood ratio is also used for the slope at every point of the ROC. 
Since we apply the likelihood ratio as a
summary index for ROC, we specify, that for our purposes the term likelihood ratio
refers only to the slope of the ROC curve at the vicinity of the
origin as in Eq.\ (\ref{defm}).

Note, that one can show that the precursor, which maximizes the
likelihood as explained in Sec.\ \ref{pre} also maximizes the $m$ and is in
this sense the optimal precursor. 
%
\subsection{Addressing the dependence on the event magnitude \label{evsize}}
We are now interested in learning how the predictability depends on the event magnitude
$\eta$ which is measured in units of the standard deviation of the time
series under study. 
Thus the event variable $Y_{n+1}$ becomes dependent on the event
magnitude
\begin{eqnarray}
Y_{n+1}(\eta) & = &\left\{  \begin{array}{ll}
1\quad&\mbox{ \small an event of magnitude $\eta$ or larger }\\
& \mbox{ occurred at time $n+1$} \\
 0\quad&\mbox{ \small no event of magnitude $\eta$ or larger}\\
&\mbox{  occurred at time $n+1$}\end{array} \right. 
\end{eqnarray}

Via Bayes' theorem the likelihood ratio can be
expressed in terms of the likelihood $L\bigl(Y_{n+1}(\eta)=1|{\mathbf
  x}_{pre}\bigr)$ and the total probability to find events
$P\bigl(Y_{n+1}(\eta)=1\bigr)$.
Inserting the technical details of the calculation of the likelihood and the
total probability (see the appendix) we can see that the likelihood
ratio depends sensitively on the joint PDF
$j(\mathbf{x}_{(n-k+1,n)},Y_{n+1}(\eta)=1)$ of pecursory variable and event.
\begin{widetext}
\begin{eqnarray}
m (Y_{n+1}(\eta),\mathbf{x}_{(n-k+1,n)})& = &\frac{\left(1 - \int_{-\infty}^{\infty} d\mathbf{x}_{(n-k+1,n)} \; j(\mathbf{x}_{(n-k+1,n)},Y_{n+1}(\eta)=1)\right)}{\int_{-\infty}^{\infty} d\mathbf{x}_{(n-k+1,n)} \; j(\mathbf{x}_{(n-k+1,n)},Y_{n+1}(\eta)=1)}
\frac{  \frac{j(\mathbf{x}_{(n-k+1,n)},Y_{n+1}(\eta)=1)}{\rho(\mathbf{x}_{(n-k+1,n)})}
}{\left(1 -
  \frac{j(\mathbf{x}_{(n-k+1,n)},Y_{n+1}(\eta)=1)}{\rho(\mathbf{x}_{(n-k+1,n)})}\right)} \label{mgeneral}, \nonumber \\
\mbox{with} &\it{} & j(\mathbf{x}_{(n-k+1,n)},Y_{n+1}(\eta)=1) =
\int_{\mathcal{M}} \; dx_{n+1}\;
j(\mathbf{x}_{(n-k+1,n)},x_{n+1}),\quad \quad  \nonumber\\
& \it{}& \mathcal{M} = \{ x_{n+1}: Y_{n+1}=1 \},\\
\mbox{and} & \it{}&\quad \rho(\mathbf{x}_{(n-k+1,n)}) =
  j(\mathbf{x}_{(n-k+1,n)},Y_{n+1}(\eta)=1) +
  j(\mathbf{x}_{(n-k+1,n)},Y_{n+1}(\eta)=0). \nonumber
\end{eqnarray}
\end{widetext}
Hence once the precursor is chosen, the dependence on the event
magnitude $\eta$ enters into the likelihood ratio, via the joint PDF of event and precursor.
Looking at the rather technical formula in Eq.\ (\ref{mgeneral}), there are
two aspects, which we find remarkable:
\begin{itemize}
\item[(I)] The slope of the ROC curve is fully characterized by the knowledge of
  the joint PDF of precursory variable and event. 
 This implies that in the framework of
  statistical predictions all kinds of (long-range) correlations which might be
  present in the time series influence the quality of the predictions
  only through their influence on the joint PDF. 
\item[(II)] The definition of the event, e.g., as a threshold crossing or an
increment does change this dependence only insofar as it enters into the
choice of the
precursor and it influences also the set on which the integrals in Eq.\ (\ref{mgeneral}) are
carried out. 
Both $Y_{n+1}(\eta)$ and the set $\mathcal{M}$ have
to be defined according to the type of events one predicts. 
 When predicting, e.g., increments $x_{n+1} - x_n \geq \eta $ via the precursory variable $x_n$, then $\mathcal{M}=[a,b]$ with
$ a(Y_{n+1}(\eta)) = x_n + \eta$ for the lower border and $ b(Y_{n+1}(\eta)) =
\infty$ for the upper border. 
In order to predict threshold crossings at $x_{n+1}$ via $x_n$ one uses  $ a(Y_{n+1}(\eta)) = \eta$, $ b(Y_{n+1}(\eta)) = \infty$. 
\end{itemize}

Exploiting Eq.\ (\ref{mgeneral}) we can hence determine the dependence of the
likelihood ratio and the ROC curve on the events magnitude $\eta$, via the
dependence of the joint PDF of the process under study.
\subsection{Constraint for increasing quality of predictions with increasing
event magnitude \label{constr}}
In order to study the dependence of the likelihood ratio on the event magnitude we are going to introduce a constraint which the likelihood and the total
probability to find events have to fulfill in order to find a better
predictability of larger (smaller) events.   
       
In order to improve the readability of the paper, we will first introduce
the following notations for %
the aposterior PDFs, the likelihood and the total probability to find events
\begin{eqnarray}
\rho_c({\mathbf x}_{(n-k+1,n)}, \eta) & = & \rho({\mathbf
  x}_{(n-k+1,n)}|Y_{n+1}(\eta)=1), \; \\ 
\rho_f ({\mathbf x}_{(n-k+1,n)}, \eta) & = &  \rho({\mathbf
  x}_{(n-k+1,n)}|Y_{n+1}(\eta)=0),\; \\
L (\eta, {\mathbf x}_{(n-k+1,n)})& = & L(Y_{n+1}(\eta)=1|{\mathbf
  x}_{(n-k+1,n)}), \;\\
P(\eta) & = & P(Y_{n+1}(\eta)=1).
\end{eqnarray}
 We can then ask for the change of the
likelihood ratio with changing event magnitude $\eta$.
\begin{equation}
\frac{\partial}{\partial \eta}\, m(Y_{n+1}(\eta),\mathbf{x}_{(n-k+1,n)})  \gtreqqless0.
\end{equation}
The derivative of the likelihood ratio is positive (negative, zero), if the following sufficient condition $c(\eta)$ is fulfilled.
\begin{widetext}
\begin{equation}
c(\eta, {\mathbf x}_{(n-k+1,n)}) = \frac{\partial}{\partial \eta}\ln L(\eta,{\mathbf x}_{(n-k+1,n)})-
\frac{\bigl(1-L(\eta,{\mathbf x}_{(n-k+1,n)}))\bigr)}{\bigl(1-P(\eta)\bigr)}\;\frac{\partial}{\partial
\eta}\ln P(\eta) \gtreqqless 0. \label{c2b}
\end{equation}
\end{widetext}
Hence one can tell for an arbitrary process, if extreme events are better
predictable, by simply  testing, if the marginal PDF of the event and the
likelihood of event and precursor fulfill Eq.\ (\ref{c2b}).

\section{Predictions of Increments in i.i.d.\ random numbers \label{num}}
In this section we test the condition $c(\eta,{\mathbf x}_{(n-k+1,n)})$ as given in Eq.\ (\ref{c2b}) for increments in Gaussian, power-law,
 and exponentially distributed i.i.d.\ random numbers.
 We thus concentrate on extreme events which consist of a sudden increase (or decrease) of the observed variable within a
few time steps. 
Examples of this kind of extreme events are the increases in
wind speed in \cite{Physa,Euromech}, but also stock market
crashes \cite{stocks,Sornette2} which consist of sudden decreases.

We define our extreme event by an increment $x_{n+1}-x_n$
exceeding a given threshold $\eta$
\begin{equation}
 x_{n+1}-x_n \geq   \eta, \label{e0}
\end{equation}
where $x_{n}$ and $x_{n+1}$ denote the observed values at two consecutive time
steps and the event magnitude $\eta$ is again measured in units of the standard deviation. 

 Since the first part of the increment $x_n$ can be used as a precursory variable, the definition of the event as an increment introduces a correlation between the event and the precursory variable
$x_n$. 
Hence the prediction of increments in random numbers provides a simple, but not unrealistic example which allows us to study the influence of the
 distribution of the underlying process on the event-magnitude dependence of the
 quality of prediction. 

In the examples which we study in this section the joint PDF of 
precursory variable and event is known and we can hence evaluate $c(\eta,x_n)$
analytically. 
A mathematical expression for a filter which selects the PDF of our
extreme events out of the PDFs of the underlying stochastic process can be
obtained through applying the  Heaviside function $ \Theta( x_{n+1} - x_{n}
-\eta)$ to the joint PDF. 
This method is described in more detail in the appendix.

Since in most cases the structure of the PDF is not known analytically, we are also
interested in evaluating
$c(\eta,x_n) $ numerically. 
In this case the approximations of the total probability and the likelihood are obtained by \lq\lq binning and counting" and
their numerical derivatives are evaluated via a
Savitzky-Golay filter \cite{Savgol,numrecipes}.
The numerical evaluation is done within $10^7$ data points. In order to check
the stability of this procedure, we evaluate $c(\eta,x_n)$ also on $20$
bootstrap samples which are generated from the original data set. 
These bootstrap samples consist of $10^7$ pairs of event and precursory variable, which were drawn randomly from the
original data set. 
Thus their PDFs are slightly different in their first and second
moment and they contain different numbers of events. 
Evaluating $c(\eta, x_n)$ on the bootstrap samples thus shows how
sensitive our numerical evaluation procedure is towards changes in the
numbers of events. This is especially important for large and therefore rare
events.

In order to check the results obtained by the evaluation of $c(\eta,x_n) $, we
compute also the corresponding ROCs analytically and numerically. 

Note that for both, the numerical evaluation of the condition and the
ROC curves, we used only
event magnitudes $\eta$ for which we found at least $1000$ events, so that the
observed effects are not due to a lack of statistics of the large events.
\subsection{Gaussian distributed random numbers \label{Gaussian}}
\begin{figure}
\epsfig{file=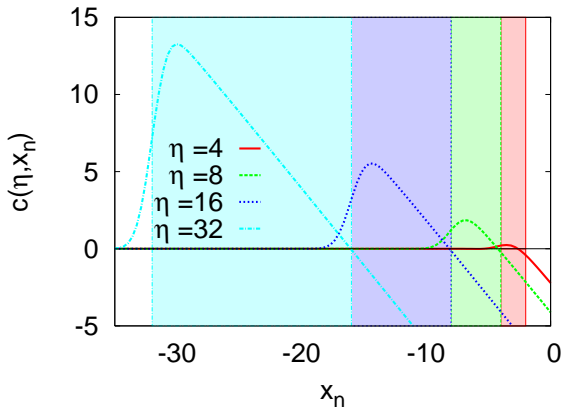,width=8cm}
\caption{The condition $c(\eta,x_n)$ for the
  Gaussian distribution as given by Eq.\ (\ref{cgauss}). 
The color shaded regions indicate the intervals $[-\sigma \eta, -\eta/2]$ for
which we can according to Eq.\ \ref{gausseins} expect
$c(\eta,x_n)$ to be positive. 
If $x_n < -\sigma \eta $, $\eta >
2\sqrt{\pi}$ and terms of the order of $\exp(-(x_n + \sigma \eta)^2)$ are
sufficiently small, the condition is also positive. 
If terms of the order of $\exp(-(x_n + \sigma \eta)^2)$ cannot be neglected one also might find
small regions in $(-\infty, -\sigma \eta]$ for which $c(\eta,x_n) <0$. 
However, the influence of these regions
is neglectable, since our
alarm interval is defined as $[-\infty,\delta]$ which implies an averaging
over several possible values of the precursory variable. \label{fig:cgauss}
} 
\end{figure}
In the first example we assume the sequence of i.i.d.\ random numbers which
form our time series to be normal distributed.
As we know from \cite{Sarah2}, increments
within Gaussian random numbers are better predictable, the more extreme
they are. 
In this section we will show that their PDFs fulfill also the
condition in Eq.\ (\ref{c2b}).
Applying the filter mechanism developed in the appendix we obtain the following expressions for the a posteriori PDFs 
\begin{equation}
\rho_c(x_n,\eta) = \frac{\exp\left(-\frac{x_{n}^2}{2\sigma^2}\right)}{2\sqrt{2\pi}\sigma
  P(\eta)}\mbox{erfc}\left(\frac{x_{n}+\sigma\eta}{\sigma\sqrt{2}}\right),\label{marga}
\end{equation}
and the likelihood
\begin{equation}
L(\eta,x_n)  =  \frac{1}{2}\erfc\left(\frac{x_n +
  \sigma\eta}{\sigma\sqrt{2}}\right). \label{gausslikely}
\end{equation}
We recall that the optimal precursor is given by the value of $x_n$ which
maximizes the likelihood. %
We refer to this special value of the variable $x_n$ by $x_{pre}$ and find for
the likelihood according to Eq.\ (\ref{gausslikely}) $x_{pre}=-\infty$. 
Thus, instead of a finite alarm volume $\delta$
here is the upper limit of the interval $[-\infty, \delta ]$. 
The total probability to find increments of magnitude $\eta$ is given by 
\begin{equation}
P(\eta) = \frac{1}{2} \erfc(\eta/2).\label{gaussptotal}
\end{equation}
\begin{figure}
\epsfig{file=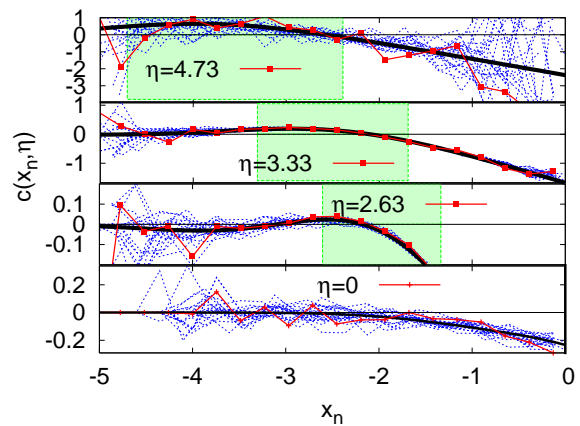 ,width=8cm}
\caption{Comparison of the numerically
  evaluated condition $c(\eta,x_n)$ for the Gaussian
distribution and the expression given by Eq.\ (\ref{cgauss}). 
The black curves denote the evaluation of the analytic result in Eq.\
  (\ref{cgauss}), the curves plotted with lines and symbols represent the numerical results
  obtained from the original data set, and the dashed lines represent the
  results obtained from the corresponding bootstrap samples. 
The gray (green in the colored plot) regions indicate the regime $-\sigma \eta
< x_n < -\sigma \eta /2$ for which $c(\eta, x_n)$ is positive in the limit
$\eta \rightarrow \infty$.
The numerical
evaluation of $c(\eta,x_n)$ was done by sampling the likelihood and the total
probability of events from $10^7$ random numbers.\label{fig:cnumgauss}}
\end{figure}
Hence the condition in Eq.\ (\ref{c2b}) reads 
\begin{eqnarray}
c(\eta,x_n) & = &-\sqrt{\frac{2}{\pi}}\frac{\exp\left(-z^2\right)}
{\erfc\left(z\right)} \quad  \nonumber\\
&\it{}& + \frac{1}{\sqrt{\pi}} \frac{\exp\left(-\frac{\eta^2}{4}\right)}{\erfc\left(\frac{\eta}{2}\right)} 
\frac{\left(1 -\frac{1}{2}\erfc\left(z\right) \right)}
{\left( 1- \frac{1}{2} \erfc\left(\frac{\eta}{2}\right)\right)}, \nonumber\\
&\it{}& \mbox{with}\quad z=\frac{x_n + \sigma \eta}{\sqrt{2}\sigma}  \label{cgauss}
\end{eqnarray}
Fig.\ \ref{fig:cgauss} illustrates this expression
and Figure \ref{fig:cnumgauss} compares it to the numerical results. 
For the ideal precursor $x_{n} = x_{pre} = -\infty $ the condition $c(\eta,x_n)$ is
---according to Eq.\ (\ref{cgauss})---
zero, since in this case,
the slope of the ROC-curve tends to infinity \cite{Sarah2} and does not react to any
variation in $\eta$.
For any finite value of the precursory variable $x_{n} <0$ we have to distinguish three regimes
of $z=(x_n + \sigma \eta)/\sqrt{2}\sigma$, namely, $z \rightarrow \infty$ or $z
\rightarrow -\infty$ and finally also the case $z=0$.

In the first case we study the
behavior of $c(\eta,x_n)$ for a fixed value of the precursory variable $-\sigma \eta <x_n$ and $\eta
\rightarrow \infty$. 
This implies that $z\rightarrow \infty$ and we can use
the asymptotic expansion for large arguments of the complementary
error function
\begin{figure}[t!!!]
\includegraphics[width=9cm]{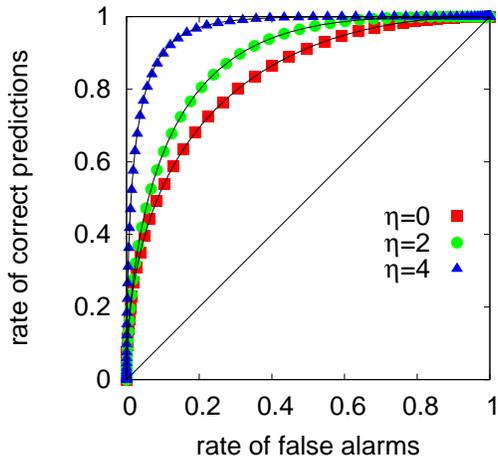} 
\caption[]{\small\label{fig:rocgauss} ROCs for Gaussian distributed i.i.d.\ random variables. 
The symbols
represent ROC curves which where made via predicting increments in $10^7$ normal
i.i.d.\ random numbers.
 The predictions were made according to the prediction
strategy described in Sec.\ \ref{pre}. The lines represent the results of evaluating
the integrals in Eqs.\ (\ref{rcor}) and (\ref{rf}) for the Gaussian case. Note that the quality of the
prediction increases with increasing event magnitude.}
\end{figure}
%
\begin{eqnarray}
\erfc(z) &\sim &  \frac{\exp(-z^2)}{\sqrt{\pi}z} \left(1 + \sum_{m=1}^{\infty}
(-1)^m \frac{1 \cdot 3 ...(2m-1)}{(2z^2)^m }\right),\nonumber\\ 
&\it{}& \left( z \rightarrow \infty, |
\mbox{arg} z | < \frac{3\pi}{4} \right)\label{erfcapprox}
\end{eqnarray}
which can be found in \cite{Abram} to obtain
\begin{eqnarray}
c(\eta, x_n) & \propto & -\frac{x_n}{\sigma} + \frac{\eta}{2},\quad -\sigma \eta < x_n <0. \label{gausseins}
\end{eqnarray} 
This expression is appropriate for $x_n > -\sigma \eta$ since the
asymptotic expansion in Eq.\ (\ref{erfcapprox}) holds only if the argument of
the complementary error function is positive. 
In this case $c(\eta, x_n)$ is larger than zero, if $x_n$ is fixed and finite and $-\sigma \eta < x_n < -\sigma \eta/2$. 

In the second case, we assume $\eta \gg 1$ to be fixed, $x_n <-\sigma \eta$ and $x_n
\rightarrow -\infty$. 
Hence we can use the expansion in Eq. (\ref{erfcapprox}) only to obtain
the asymptotic behavior of the dependence on $\eta$ and not for the dependence
on $z$.
An asymptotic expression of $c(\eta, x_n)$ hence reads 
\begin{eqnarray}
c(\eta, x_n) & \propto & \frac{\eta}{2\left(1 - \frac{1}{2}
  \erfc\left(\eta/2\right) \right)} \left(\frac{\erf(z)}{\sqrt{pi}} +
  \frac{\eta}{2} \right) \nonumber\\
 &\it{}& - \mathcal{O}\left(\exp(-z^2) \right), \quad x_n < -\sigma
  \eta. \label{gausszwei}
\end{eqnarray}
Since $\erf(z)$ tends to minus unity as $z \rightarrow -\infty$ the expression
in Eq. (\ref{gausszwei}) is positive if $\eta > 2\sqrt{\pi}$ and if we can
assume the squared exponential term to be sufficiently small.
If the later assumption is not fullfilled one might observe some regions of
intermediate values of $-\infty< x_n < -\sigma \eta$, for which $c(\eta,x)$ is
negative.

However the ROC curves in Fig.\ \ref{fig:rocgauss} suggest that the influence
of these regions is sufficiently small, if the alarm volume is chosen to be $[-\infty, \delta ]$. 
We can understand this effect, if we keep in mind that we use the interval
$[-\infty, \delta]$ as an alarm volumen. 
Hence we can expect that the influence of the regions, where $c(\eta, x_n)$ is negative, is suppressed since we average over many different values of $x_n$ and the condition is positive as $x_n \rightarrow -\infty$. 
(Positive is meant here
in the sense, that $c(\eta,x_n)$ approaches the value zero for $x_n = -\infty$
from small positive numbers.)

In the third case, for $x_n = -\sigma \eta$ and hence $z=0$ we find that $c(\eta,x_n)$ is
positive if $\eta > 2\sqrt{\frac{2}{\pi}} \left(1 - \frac{1}{2}\erfc(\eta/2) \right)$.

In total we can expect larger increments in Gaussian random numbers to be easier
to predict the larger they are. 
The ROCs in Fig.\ \ref{fig:rocgauss} support these results.


\subsection{Symmetrized exponential distributed random variables
  \label{symexpo}}
The PDF of the symmetrized exponential reads 
\begin{eqnarray}
\rho(x) & = & \frac{\lambda}{2} \exp(-\lambda |x_n|) = \left\{ \begin{array}{l@{\quad:\quad}l} \frac{\lambda}{2} \exp(-\lambda x_n) & x_n
>0, \\ \lambda/2 &  x_n =0,\\ \frac{\lambda}{2} \exp(\lambda x_n) & x_n
<0, \end{array} \right.  \nonumber
\end{eqnarray}
with $\mu =0$, $\sigma = \sqrt{2}/\lambda$. 

Applying the filtering mechanism
according to the appendix we find the joint PDFs of precursory variable and event 
\begin{widetext}
\begin{eqnarray}
j(x_n,(Y_{n+1}(\eta)=1)) & = & \left\{ \begin{array}{l@{\quad :\quad}l} 
\frac{\lambda}{4}\exp(-\sqrt{2} \eta -2\lambda x_n) & x_n >0,\\
\frac{\lambda}{4}\exp(-\sqrt{2}\eta) & -\eta <x_n <0,\\
 \frac{\lambda}{2} \left(\exp(\lambda x_n) -\frac{1}{4}\exp\left(\sqrt{2}\eta 
     +\lambda 2x_n \right) \right) & x_n < - \eta <0,\\  
\end{array} \right. 
\end{eqnarray}
\end{widetext} 
\begin{widetext}
the aposterior probabilities,
\begin{eqnarray}
\rho_c(x_n, \eta, \lambda) & = & \left\{\begin{array}{l@{\quad :\quad}l} 
  \frac{\lambda}{(2 + \sqrt{2}\eta}\exp( -2\lambda x_n) & x_n >0,\\
\frac{\lambda}{(2+\sqrt{2}\eta}& -\eta <x_n <0,\\
 \frac{\lambda}{(2+ \sqrt{2}\eta} \left(2\exp(\sqrt{2}\eta + \lambda x_n) -\exp\left(2\sqrt{2}\eta 
     +2\lambda x_n \right) \right) & x_n < - \eta <0,\\  
\end{array} \right. \label{symexpapostc}\\
\rho_f(x_n,\eta, \lambda) & = & \left\{\begin{array}{l@{\quad :\quad}l} 
  \frac{\lambda}{2}\exp( -\lambda x_n) \frac{\left(1 - \frac{1}{2}
  \exp(-\lambda x_n - \sqrt{2}\eta) \right)}{\left(1 -
  \frac{1}{2}\left( 1 + \frac{\eta}{2}\right)  \exp(-\sqrt{2}\eta) \right)} & x_n >0,\\
 \frac{\lambda}{2}\exp( \lambda x_n) \frac{\left(1 - \frac{1}{2}
  \exp(-\lambda x_n - \sqrt{2}\eta) \right)}{\left(1 -
  \frac{1}{2}\left( 1 + \frac{\eta}{2}\right)  \exp(-\sqrt{2}\eta) \right)} & -\eta <x_n <0,\\
 \frac{\lambda}{4} \frac{\exp(2\lambda x_n + \sqrt{2}\eta)}{\left(1 -
  \frac{1}{2}\left( 1 + \frac{\eta}{2}\right)  \exp(-\sqrt{2}\eta) \right)} & x_n < - \eta <0,\\  
\end{array} \right. \label{symexpapostf}
\end{eqnarray}
\end{widetext} 
\begin{widetext}
the likelihood 
\begin{eqnarray}
 L(\eta,x_n, \lambda) & = & \left\{ \begin{array}{l@{\quad :\quad}l} 
\frac{1}{2}\exp(-\sqrt{2} \eta -\lambda x_n) & x_n >0,\\
\frac{1}{2}\exp(-\sqrt{2}\eta - \lambda x_n) & -\eta <x_n <0,\\
 1 -\frac{1}{2}\exp(\sqrt{2}\eta  +\lambda x_n ) & x_n < - \eta <0,\\  \end{array} \right. \label{symexplike}
\end{eqnarray}
\end{widetext}
and the total probability to find events of magnitude $\eta$
\begin{eqnarray}
P(\eta) & = & \left\{ \begin{array}{l@{\quad :\quad}l} 
\frac{1}{8}\exp(-\sqrt{2} \eta) & x_n >0,\\
\frac{\sqrt{2}}{4}\eta \exp(-\sqrt{2}\eta) & -\eta <x_n <0,\\
\frac{3}{8}\exp(-\sqrt{2}\eta) & x_n < - \eta <0.  \end{array}\right. \quad 
\end{eqnarray}
If we are not interested in the range of the precursory variable, the total probability
to find events is given by 
\begin{equation}
P(\eta)  =  \frac{1}{2}\exp(-\sqrt{2}\eta) \left(1 + \frac{\eta}{\sqrt{2}} \right).
\end{equation}
Hence the condition $c(\eta, x_n, \lambda )$ reads
\begin{widetext}
%
\begin{figure*}[t!!!]
\parbox{18cm}{
\parbox{8.5cm}{
\epsfig{file=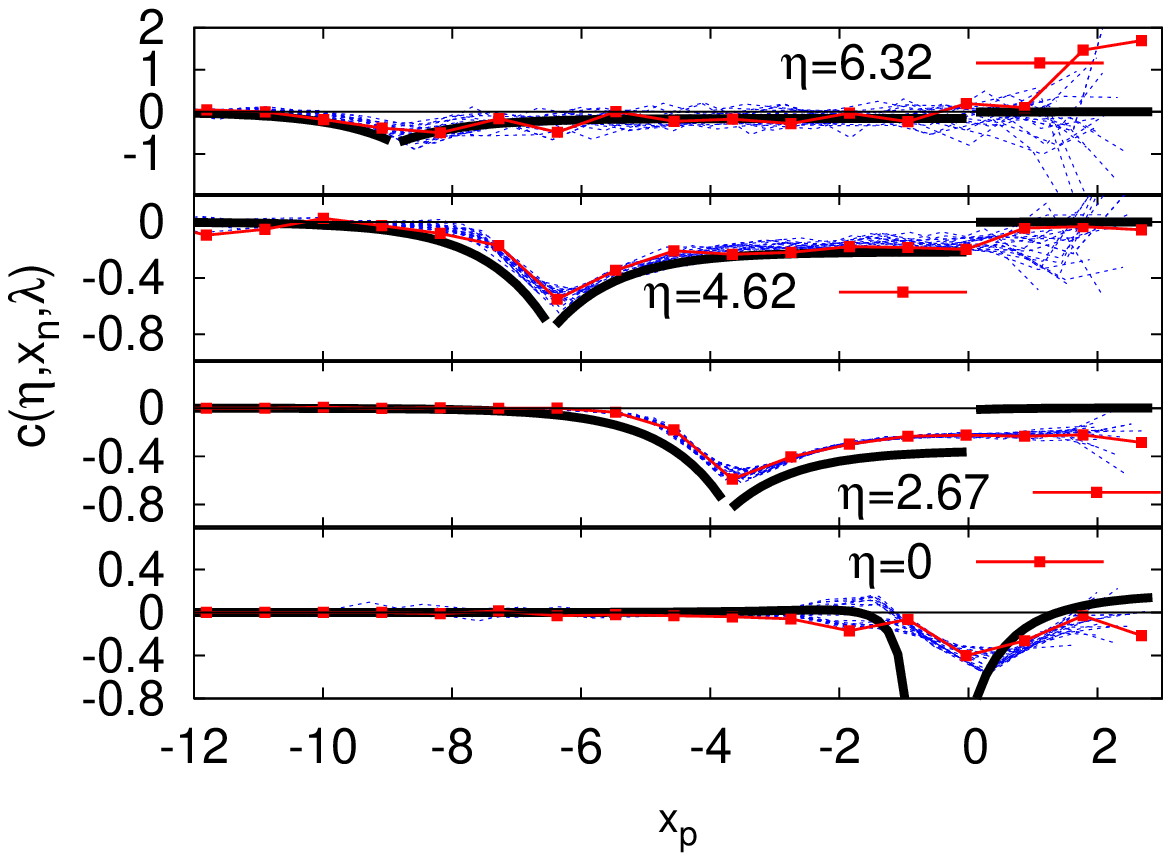, width=8cm}
\caption{The numerically and analytically evaluated condition
  for the symmetrized exponential. The black line is the result of the
  analytical evaluation according to  Eq.\ (\ref{csymexp}), the
  curves plotted with lines and symbols represent the numerical results
  obtained from the original data set, and the dashed lines represent the
  results obtained from the corresponding bootstrap samples. Note that for
  small values of $x_n$ the condition $c(\eta, x_n, \lambda)$ is for all values of $\eta$ close
  to zero.\label{fig:csymexp}}}
\hspace*{0.5cm}
\parbox{8.5cm}{
\epsfig{file=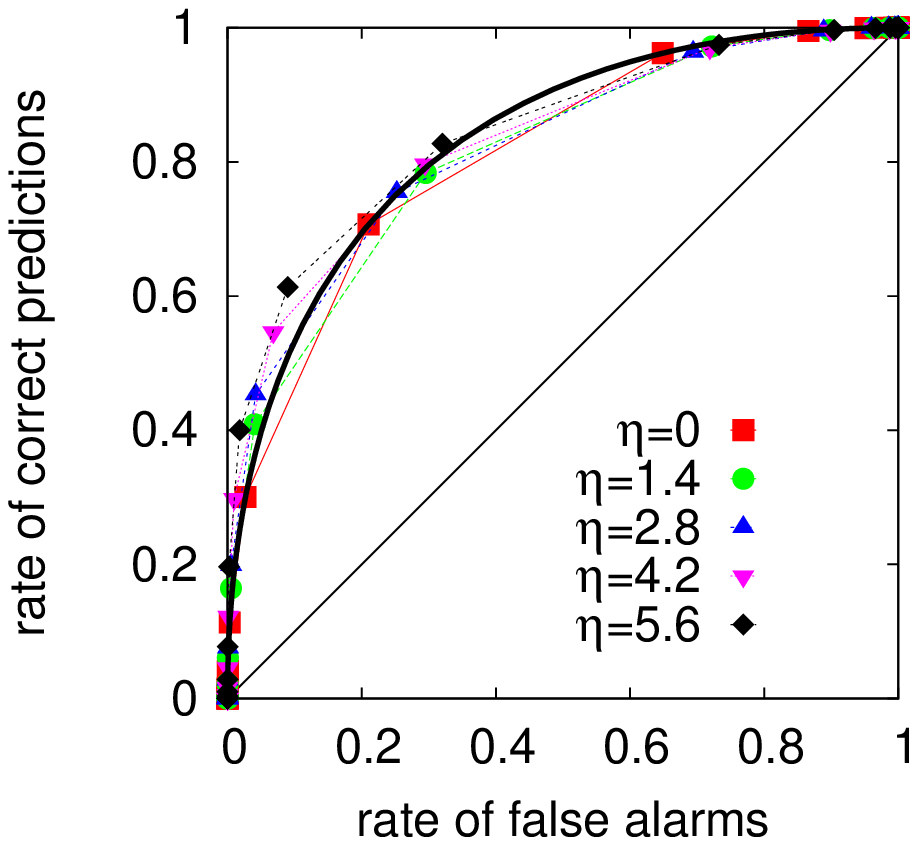,width=8cm}
\caption{The ROCs for symmetrically exponentially
  distributed i.i.d.\ random numbers show no significant dependence on the
  event magnitude.  The ROC curves were made via
  predicting increments in $10^7$ normal i.i.d.\ random numbers and the predictions were made according to the prediction
strategy described in Sec.\ \ref{pre}. The black
  line indicates the analytically evaluated ROC curve for
  $\eta=0$. \label{fig:rocsymexp}}
}
}
\end{figure*}
\begin{eqnarray}
c(\eta, x_n,\lambda) & = & \left\{\begin{array}{l}
-\sqrt{2}\left(1 - \frac{\left(1 - \frac{1}{2} \exp(-\sqrt{2}\eta - \lambda x_n)\right)}{\left(1 - \frac{1}{8}\exp(-\sqrt{2}\eta)\right)}\right), \quad x_n >0,\\
\\
-\sqrt{2} + \frac{(1 -\sqrt{2}\eta)}{\eta} \frac{\left(1- \frac{1}{2}\exp(-\sqrt{2}\eta
  -\lambda x_n)\right)}{\left(1 -
  \frac{\sqrt{2}}{4}\eta \exp(-\sqrt{2}\eta)\right)}, \quad -\eta<x_n<0,\\
\\
-\frac{1}{\sqrt{2}}\exp(\lambda x_n + \sqrt{2}\eta)\left(\frac{1}{1 -
  \frac{1}{2}\exp(\sqrt{2}\eta + \lambda x_n)} + \frac{1}{1 -
  \frac{3}{8}\exp(-\sqrt{2}\eta)} \right),\quad x_n < - \eta. 
 \end{array}\right.  \label{csymexp}
\end{eqnarray} 
\end{widetext}

Figure \ref{fig:csymexp} compares the
results of the numerical evaluation of the condition and the analytical
expression given by Eq.\ (\ref{csymexp}). Since most precursors of large
increments can be found among negative values, the numerical evaluation of
$c(\eta,x_n\lambda)$  becomes worse for positive values of $x_n$, since in
this limit the likelihood is not very well sampled from the data. This leads
also to the wide spread of the bootstrap samples in this region. 

Figure \ref{fig:csymexp} shows that in the vicinity of the smallest value of the data set, the condition
$c(\eta,x_n,\lambda)$ is zero. As we approach larger values of $\eta$,
$c(\eta,x_n,\lambda)$ approaches zero in the whole range of data values. That
is why we would expect to see no influence of the event magnitude on the quality of predictions in the exponential case. 

The ROC curves in Fig. \ref{fig:rocsymexp} support these results. The
numerical ROC curves were made via predicting increments in $10^7$ normal
i.i.d.\ random numbers according to the prediction
strategy described in Sec. \ref{pre}.
The precursor for the ROC-curves is chosen as the maximum of the likelihood
according to Eq.\ (\ref{symexplike}), i.e., $x_{pre}=-\infty$, so that the
alarm interval is $[\infty,\delta]$.
In summary there is no significant dependence on the event magnitude for the prediction
of increments in a sequence of symmetrical exponential distributed random numbers.
\subsection{Pareto distributed random variables \label{powl}}
\begin{figure}[t!!!]
\includegraphics[width=6cm,angle=-90]{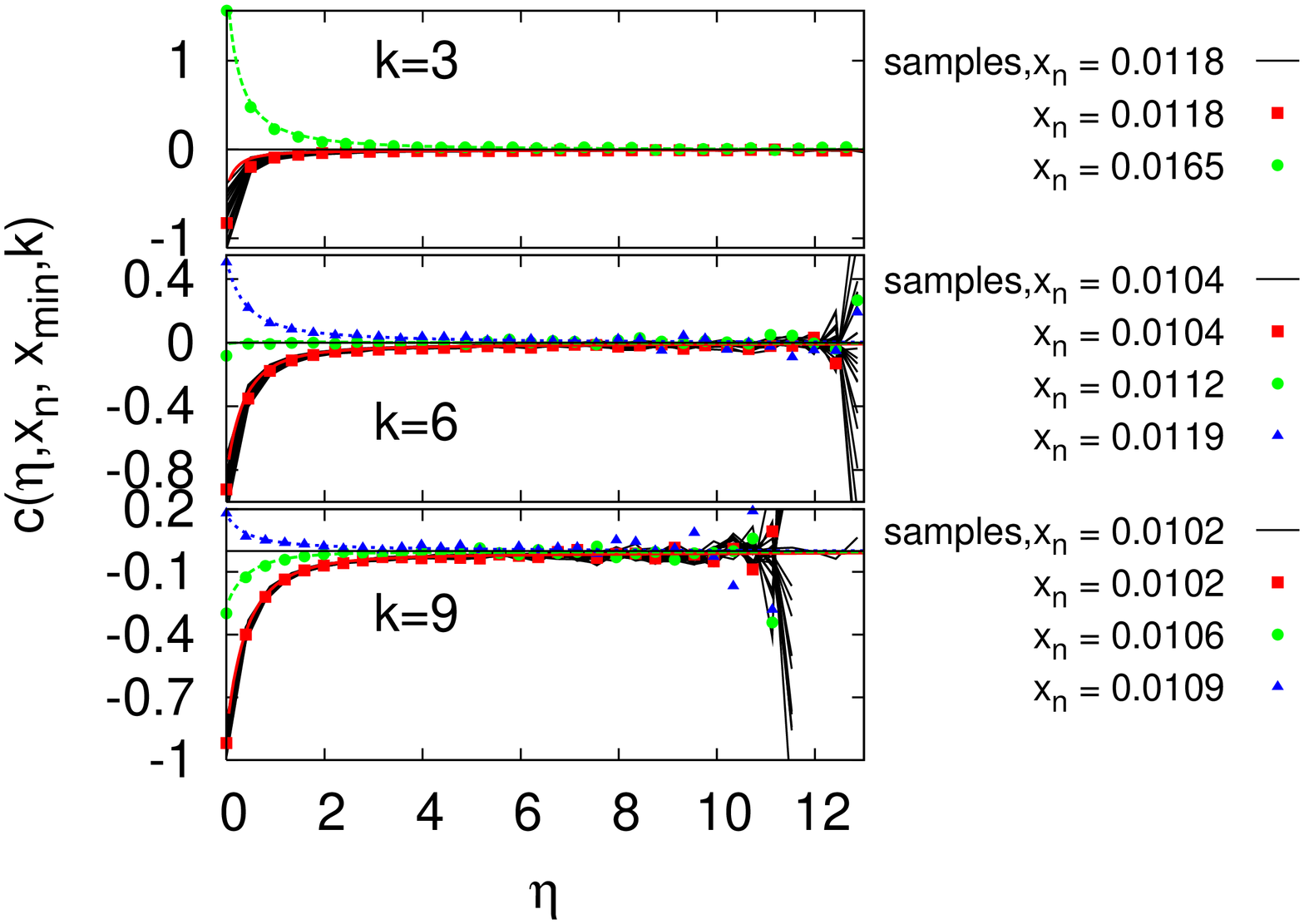} 
\caption[]{\small The condition $c(\eta,x_n,x_{min},k)$ for the power-law distribution with lower endpoint
$x_{min} =0.01$ are plotted for constant values of the precursory variable
$x_{n}$. The symbols represent the results of the numerical evaluation of
$c(\eta,x_n,x_{min},k)$, the gray (colored) lines denote the analytic results,
and the black lines denote the result for the corresponding bootstrap samples
and the optimal precursor.
For the \lq\lq ideal" precursor $x_{n}=x_{min}=0.01$ all
values of $c(\eta,k, 0.01)$ are negative. Hence one should expect smaller
events to be better predictable. However, this effect is sensitive of the
choice of the precursor. \label{fig:powerlawcompare}} 
\end{figure}
We investigate the Pareto distribution as an example for power-law
distributions. 
The PDF of the Pareto distribution is defined as \cite{Feller}
\begin{equation}
\rho(x) = kx_{min}^k \, x^{-(k+1)}
\end{equation}
for  $x \in [x_{min}, \infty)$ with the exponent $k\geq3$, the lower endpoint
$x_{min} > 0$, and variance $\sigma = \frac{x_{min}}{k-1}\sqrt{\frac{k}{k-2}}$.
Filtering for increments of magnitude $\eta$ 
we find the following conditional PDFs of the increments:
\begin{eqnarray}
\rho_c ( x_n, \eta, x_{min},k)& = & \frac{k x_{min} ^{2k}}{ x_n^{k+1} \left(x_n +
 \frac{x_{min}}{k+1}\sqrt{\frac{k}{k-2}}\, \eta \right)^k P(\eta,k)}
\label{paretoapost}\nonumber\\
\\
\rho_f (x_n, \eta, x_{min},k)& = & \frac{kx_{min}^k}{x_n^{k+1}} \frac{\left(1 -
  \left(\frac{x_{min}}{x_n +  \frac{x_{min}}{k+1}\sqrt{\frac{k}{k-2}} \,\eta
  }\right)^k\right)}{1 - P(\eta,k)}\nonumber\\
\\
L (\eta, x_n, x_{min},k)& = &\left( \frac{x_{min}}{x_n +
  \frac{x_{min}}{k+1} \sqrt{\frac{k}{k-2} \, \eta}}\right)^k .
\label{paretolikeli}
\end{eqnarray}
Within the range $(x_{min}, \infty)$ the likelihood has no well defined
maximum. However, since the likelihood is a monotonously decreasing
function, we use the lower endpoint $x_{min}$ as a precursor.
The total probability to find events of magnitude $\eta$ is given by
\begin{eqnarray}
P(\eta,k)& = & \frac{1}{2} \,{}_{2}F_1
\left(k, 2k, 2k+1,- \frac{\eta}{(k+1)}\sqrt{\frac{k}{k-2}}\right), \nonumber \\\label{paretototal}
\end{eqnarray}
where $ {}_{2}F_1(a,b,c,x) $ denotes  the hypergeometric function $ {p}_{2}F_q(a,b,c,x)$ with
$p=2$, $q=1$.

\begin{figure}[t!!!]
\includegraphics[width=6cm, angle= -90]{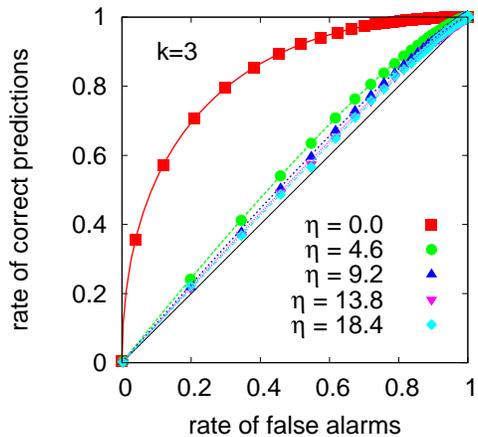} 
\caption[]{\small\label{fig:powlroc3}ROC-plot for the power-law distribution with $k=3$ and $x_{min} =0.01$. The
symbols show the numerical results and the lines indicate the analytically
calculated ROC curves.
The ROC curves were made via predicting increments in $10^7$ Pareto
distributed i.i.d.\ random numbers. The predictions were made according to
the prediction strategy described in Sec.\ \ref{pre}. Note that we tested only
event magnitudes $\eta$, for which we found at least $1000$ events, so that the
effects we observe are not due to a lack of statistics of the large events. The ROC curves display
that in Pareto distributed i.i.d.\ random numbers with the lower endpoint
$x_{min}=0.01$ smaller events are better predictable and that large events are
very hard to predict.}
\end{figure}


Using
\begin{eqnarray}
\ \frac{\partial P(\eta, k)}{\partial \eta} & = & \frac{k}{\eta} \left(
\frac{1}{\left(1 + \frac{\eta}{k+1}\sqrt{\frac{k}{k-2}} \right)^k} - 2P(\eta,k)\right) \nonumber\\
\end{eqnarray}
and inserting the expressions (\ref{paretolikeli}) and (\ref{paretoapost}) for the components of
$c(\eta,x_n, x_{min},k)$ we can obtain an explicit analytic expression for
the condition. In Fig.\ \ref{fig:powerlawcompare} we evaluate this expression using {\it
  Mathematica} and compare it with the results of an empirical evaluation on
the data set of $10^7$ i.i.d.\ random numbers.

\begin{figure}[t!!!]
\includegraphics[width=6cm, angle= -90]{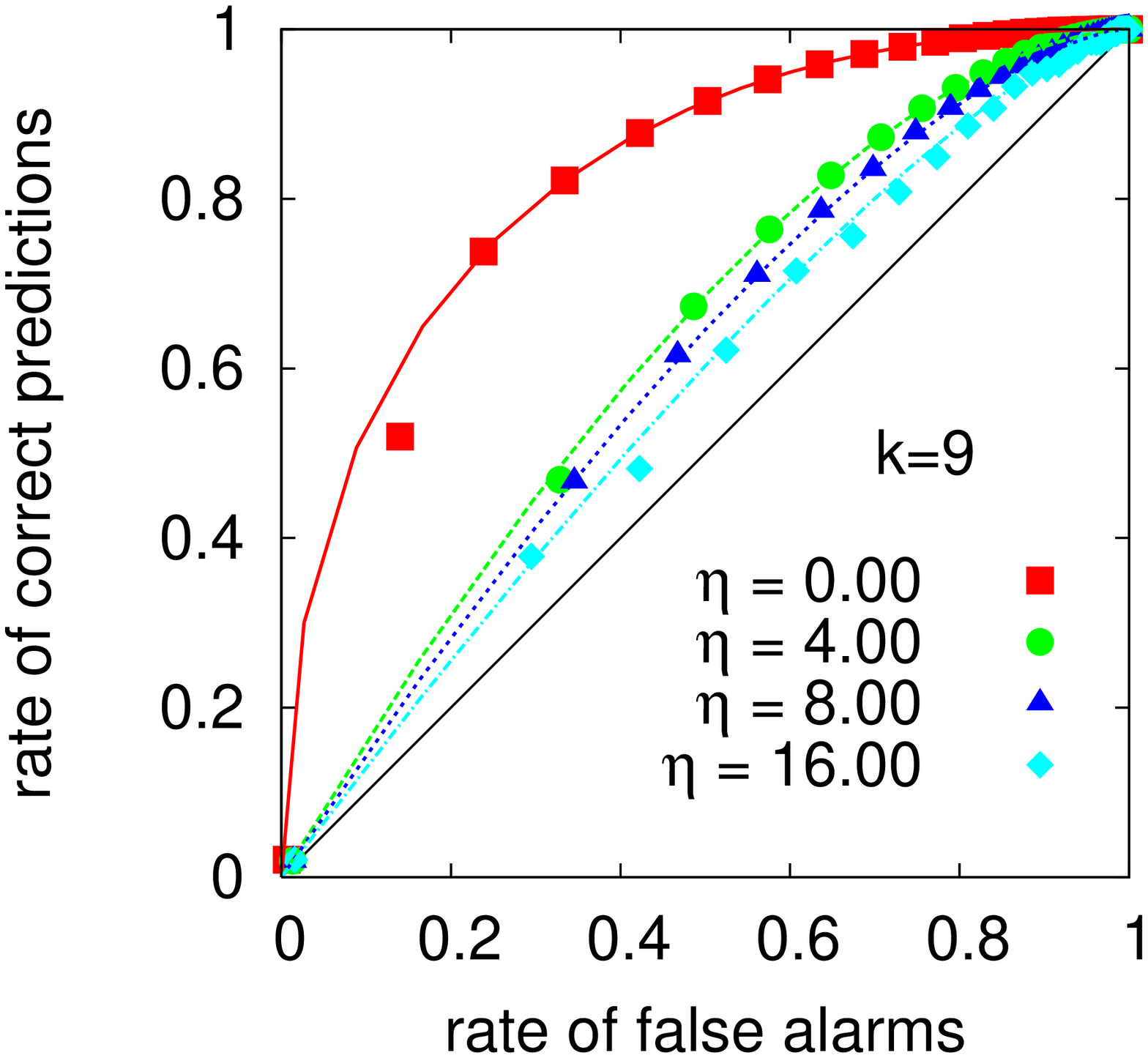} 
\caption[]{\small\label{fig:powlroc9}ROC-plot for the power-law distribution with $k=9$ and $x_{min} =0.01$.
The symbols show the numerical results and the lines indicate the analytically
calculated ROC curves.
The ROC curves where made via predicting increments in $10^7$ Pareto distributed
i.i.d.\ random numbers and the predictions were made according to the prediction
strategy described in Sec.\ \ref{pre}.
The ROC-curves display that in Pareto distributed i.i.d.\ random numbers smaller events are better
predictable and large events are especially hard to predict.}
\end{figure}

Figure (\ref{fig:powerlawcompare}) displays that the value
of $c(\eta,x_n, x_{min},k)$ depends sensitively on the choice of the
precursor.
 For the ideal precursor $x_{pre}=x_{min}$ all values of $c(\eta,k, x_{min})$ are negative. 
Hence one should in this case expect smaller
events to be better predictable. 
The corresponding ROC curves in Figs.\
\ref{fig:powlroc3}, 
and \ref{fig:powlroc9} verify this
statement of $c(\eta,x_n, x_{min},k)$.

In summary we find that larger events in Pareto distributed i.~i.~d.\ random
numbers are harder to predict the larger they are. This is an admittedly
unfortunate result, since extremely large events occur much more frequently in
power-law distributed processes than in Gaussian distributed processes. Hence,
their prediction would be highly desirable.

\section{Increments in Free Jet Data \label{freejet}}
\begin{figure}
\epsfig{file=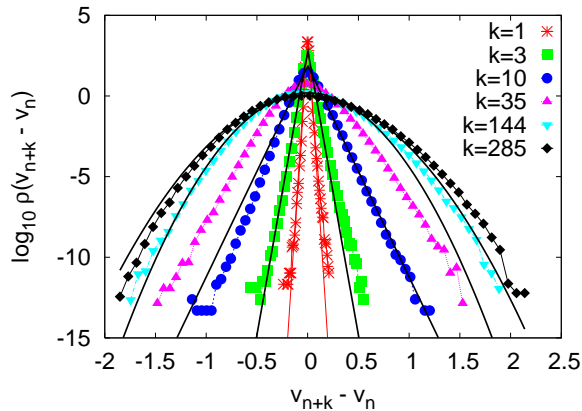,width=8cm}
\caption{\label{fjhisto}PDF of the increments $a_{n,k}=v_{n+k}-v_n$ with $k=
  1,3,10,35,144,285$. The black lines correspond to Gaussian and exponential
  PDFs with appropriate values for the standard deviation or the coefficient $\lambda$.}
\end{figure}

In this section, we apply the method of statistical inference to predict
acceleration increments in free jet data. 
Therefore we use a data set of
$1.25\times 10^7$ samples of the local velocity measured in the turbulent
region of a round free jet \cite{Peinke}. 
The data were sampled by a hot-wire
measurement in the central region of an air into air free jet. 
One can then calculate the PDF of velocity increments  $a_{n,k}=v_{n+k} -v_n$, where $v_n$ and
$v_{n+k}$ are the velocities measured at time step $n$ and $n+k$. The Taylor
hypothesis allows one to relate the time-resolution to a spatial
resolution \cite{Peinke}. 
One observes that for large values of $k$ the PDF of increments is essentially indistinguishable
from a Gaussian, whereas for small $k$, the PDF develops approximately
exponential wings \cite{vanAtta, Frisch1990, Frisch1995}. 
Fig.\ \ref{fjhisto} illustrates this effect using the data set under study.
\begin{figure*}[t!!!]
\centerline{
\epsfig{file=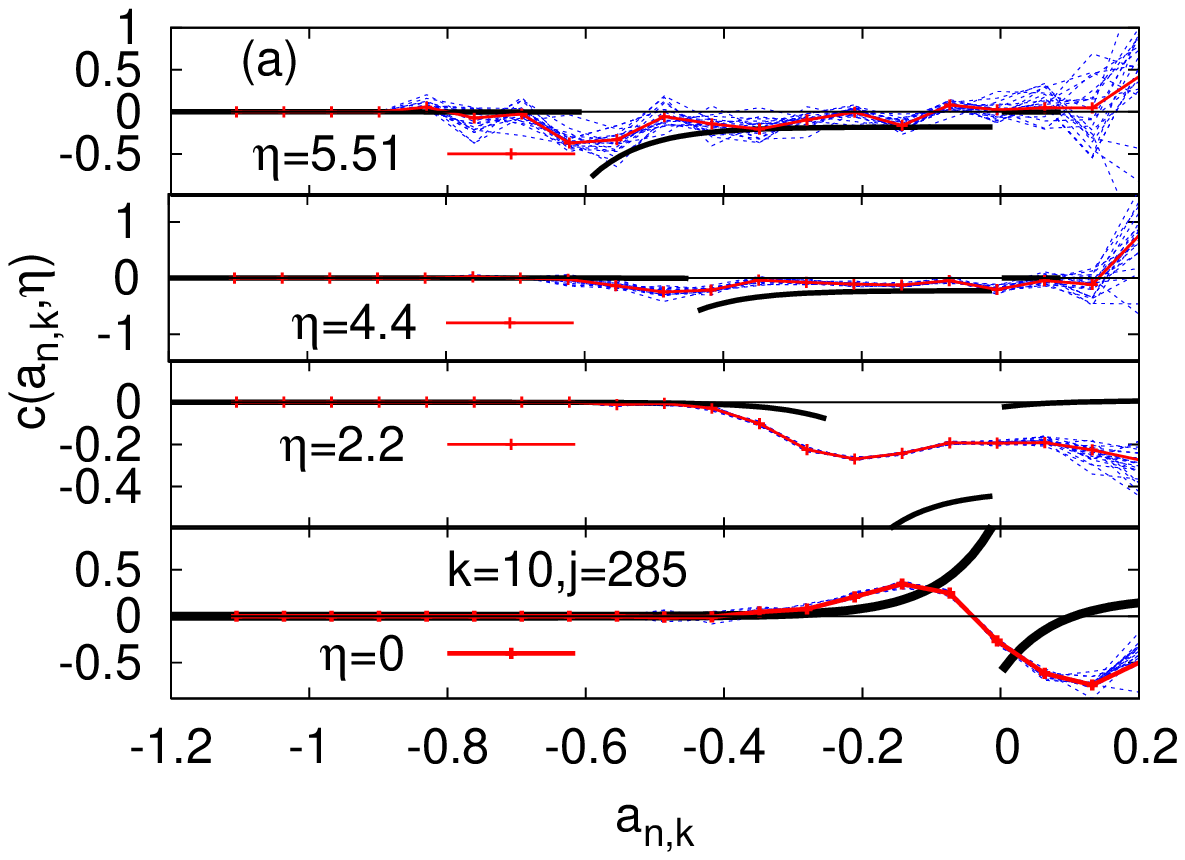,width=8cm} \hspace{0.01cm}
\epsfig{file=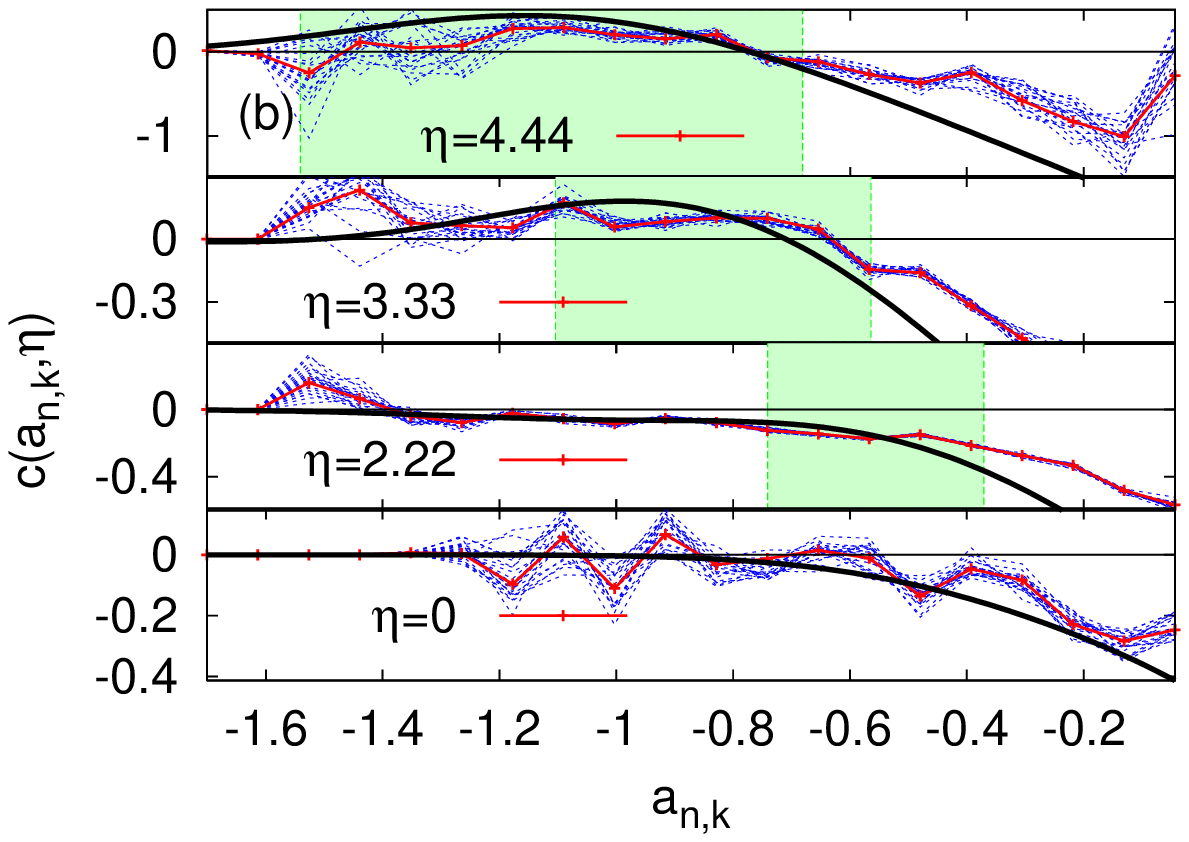,width=8cm}}
\caption[]{\label{fjcondi} Transition from the exponential regime
  (a) to the Gaussian regime (b) characterized via the numerical evaluation of
  $c(x_n,\eta)$. The black line corresponds to the analytic results for the
  Gaussian and the exponential PDF, fitted to the PDFs of the
  increments, as it is shown in Fig.\ \ref{fjhisto}. For larger values of $\eta$ the main features of
$c(x_n,\eta)$ for the exponential and the Gaussian case as described in
Sec.\ \ref{Gaussian} and \ref{symexpo} are reproducable. For larger values of $\eta$
we find that if $-\sigma \eta < a_{n,k} -\sigma \eta/2 $ $c(a_n,\eta)$ is either
larger than zero in the asymptotically Gaussian case ($k=144$) or equal to zero in the
asymptotically exponential case ($k=10$).}
\end{figure*}
Thus the incremental data sets $a_{n,k}$ provides us with the opportunity to test the
results for statistical predictions within Gaussian and exponential
distributed i.i.d.\ random numbers on a data set, which exhibits
correlated structures. 

We are now interested in predicting increments of the acceleration
$a_{n+j,k} - a_{n,k} \geq \eta$  in the incremental data sets  $a_{n,k}=v_{n+k}
-v_n$. 
In the following we concentrate on the incremental data set $a_{n,10}$, which
has an asymptotically exponential PDF and the data set $a_{n,144}$, which has
an asymptotically Gaussian PDF. 
Furthermore we focus on increments between
 relatively large time steps, i.e., $j=285$, so that the short-range persistence of the
process does not prevent large events from occuring.
As in the previous
sections we are hence exploiting the statistical properties of the time series
to make predictions, rather than the dynamical properties.

We can now use the evaluation algorithm which was tested on the previous
examples to evaluate the condition for these data sets. 
The results are shown in Fig.\ \ref{fjcondi}.
We find that at least for larger values of $\eta$ the main features of
$c(x_n,\eta)$ for the exponential and the Gaussian case as described in
Sec. \ref{Gaussian} and \ref{symexpo} are also present in the free jet
data. 
For larger values of $\eta$, $c(a_{n,k},\eta)$ is either
larger than zero in the Gaussian case ($k=144$) or equal to zero in the
exponential case ($k=10$) in the region of interesting precursory variables, i.e., small
values of $a_{n,k}$.
\begin{figure*}
\centerline{
\hspace{2cm}
\epsfig{file=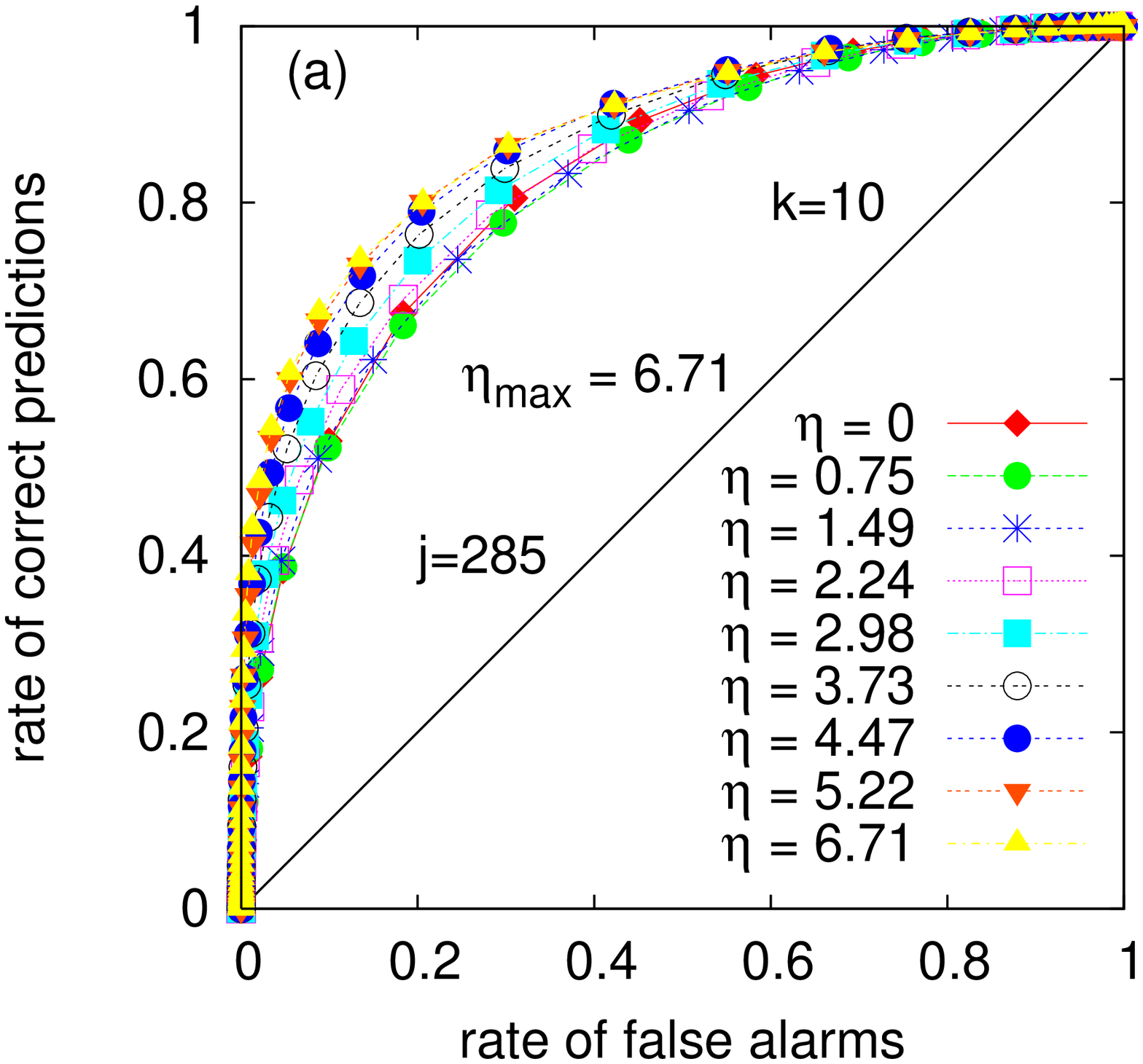,width=6cm, angle=-90}
\epsfig{file=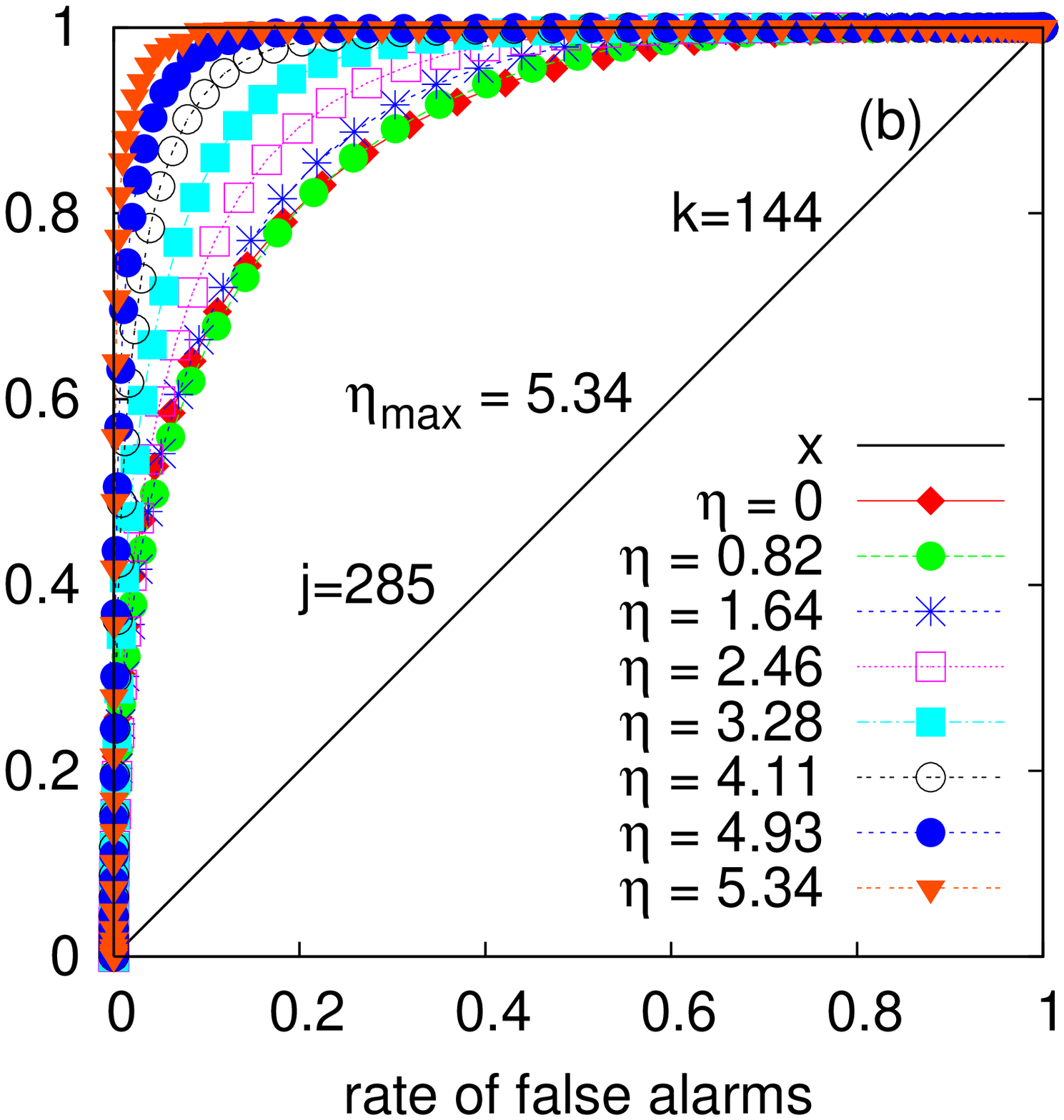,width=6cm, angle= -90}
}
\caption{\label{fjrocs}Transition from exponential ROC curves (a)  to Gaussian
  ROC curves (b). In the exponential case ($k=10$),
shown in (a) the ROC curves for different event magnitude
$\eta$ are almost the same, although the range of $\eta$ is larger ($\eta \in
(0,6.71)$) than in the Gaussian case shown in (b). For $k=144$ the ROC curves are further
apart, which corresponds to the results for Gaussian ROC curves (see
Sec.\ \ref{Gaussian}) }
\end{figure*}

However, the presence of the exponential and the Gaussian distributions is more
prominent in the corresponding ROC curves. For the free jet data set, the
predictions were made with an algorithm similar to the one described in
Sec. \ref{pre}. Instead of a specific precursory
structure, which corresponds to the maximum of the likelihood, we use here a
threshold of the likelihood as a precursor. In this setting we give an alarm
for an extreme event, whenever the likelihood that an extreme event follows an
observation is larger than a given threshold value.

In the exponential case ($k=10$)
shown in Fig.\ \ref{fjrocs}(a) the ROC curves for different event magnitude
$\eta$ almost coincide, although the range of $\eta$ is larger ($\eta \in
(0,6.71)$) than in the Gaussian case shown in Fig.\ \ref{fjrocs} (b). For $k=144$ the ROC curves are further
apart, which corresponds to the results of Secs. \ref{Gaussian} and \ref{symexpo}. 

This example of the free jet data set shows that the specific dependence of
the ROC curve on the event magnitude can also in the case of correlated data
sets be characterized by the PDF of the underlying process. 
\section{Conclusions \label{conclusions}}
We study the magnitude dependence of the quality of predictions for increments in a
time series which consists in sequences of i.i.d.\ random numbers and in
acceleration increments measured in a free jet flow. 
Using the first part of
the increment $x_n$ as a
precursory variable we predict large increments $x_{n+1} -x_n$ via statistical
considerations. 
In order to measure the quality of the predictions we use ROC curves. 
Furthermore we introduce a quantitative criterion which can determine
whether larger or smaller events are better predictable. This criterion is
tested for time series of Gaussian, exponential and Pareto i.i.d.\ random
variables and for the increments of the acceleration in the free jet flow. The
results obtained from the criterion comply nicely with the corresponding ROC-curves. 
Note that for both, the numerical evaluation of the condition and the
ROC-plots, we used only event magnitudes $\eta$ for which we found at least $1000$
events, so that the observed effects are not due to a lack of statistics of
the large events.

In the sequence of Gaussian i.i.d.\ random numbers, we find that large
increments are better predictable the larger they are. In the Pareto
distributed time series we observe that in slowly decaying power laws larger
events are harder to predict, the larger they are. We find no significant
dependence on the event-magnitude for the sequence of exponentially i.i.d.\ random numbers.

While the condition can be easily evaluated analytically, it is not that easy 
to compute numerically from observed data, since the
calculation implies evaluating the derivatives of numerically obtained
distributions. 
Using Savitzky-Golay filters improved the results, but
especially in the limit of larger events, where the distributions are
difficult to sample, one cannot trust the results of the numerically
evaluated criterion. 
However, it is still possible to apply the criterion by
fitting a PDF to the distribution of the underlying process and then evaluate
the criterion analytically.

Although the magnitude dependence of the quality of predictions was observed in
different contexts and for different measures of predictability, in this
contribution only ROC curves were used. In order to exclude the possibility
that the effect is specific to the ROC curve, future works should also include
other measures of predictability. 

Reviewing the results for the Gaussian case and the slowly decaying power law
from a philosophic point of view one can conclude that nature allows us to
predict large events from the most frequently occuring distribution
easily. However in Gaussian distributions very large events are rare and
therefore less likely to cause damage. Whereas in the less frequently occurring
distributions with heavy power-law tails, large events are especially hard to predict. 
Therefore one can assume, that rare large impact events of processes with power-law
distributions will remain unpredictable, although their prediction would be highly desirable.

\acknowledgements
We thank J. Peinke and his group for supplying us with the free jet data.
\begin{appendix}

\section{Obtaining the analytic expression for the likelihood, the joint and the aposterior PDFs for increments in stochastical processes \label{appA}}
An analytic expression for a filter which selects the PDF of our
extreme increments $x_{n+1} - x_n \geq d$ out of the PDFs of the underlying stochastic process can be
obtained through the  Heaviside function $ \Theta( x_{n+1} - x_{n}-d)$. 
(Note that $d$ is not scaled by the
standard deviation, i.e., $d =\sigma\eta$.) 
This filter is then applied to the joint PDF $j(x_0, x_1, ...,x_{n-k+1}, x_{n-k+2},
..., x_{n})$ of a stochastic process or to be more precise to the likelihood
$L(x_{n+1}|x_0, x_1, ...,x_{n-k+1}, x_{n-k+2}, ..., x_{n})$ that the $n+1$
step follows the previously obtained values. If we condition only on the last
$k$ values, we neglect the dependence on the past.
The likelihood that an event Y(d)=1 follows in the $n+1$th step can then be
obtained by multiplication with $\Theta( x_{n+1} - x_{n}-d)$.
\begin{widetext} 
\begin{eqnarray}
L(Y_{n+1}(d)=1| \mathbf{x}_{(n-k+1,n)}) & = & \Theta( x_{n+k} - x_{n}-d) L(x_{n+1}|\mathbf{x}_{(k,n)} ),
\end{eqnarray}

where $\mathbf{x}_{(n-k+1,n)}= (x_{n-k+1},x_{n-k+2}, ...,x_n) $  as defined in Sec.\ \ref{pre}.
\end{widetext}
If the resulting
expression is nonzero, the condition of the extreme event in Eq.\ (\ref{e0}) is
fulfilled and for $x_{n+1}$ and $x_{n}$ the following relation holds:
\begin{eqnarray}
x_{n+1} & = & x_{n} + d + \gamma \label{gammadef} \quad (\gamma \in  \mathbb R,
\gamma \geq 0) \quad. \label{gamma0}
\end{eqnarray}
Hence it is possible to express the likelihood in terms of $x_{n}$, which is a part of the precursory structure. We can use the integral representation of the Heaviside function with appropriate substitutions to obtain
\begin{widetext}
\begin{eqnarray}
L(Y_{n+1}(d)=1|\mathbf{x}_{(n-k+1,n)}) &=& \int_{0}^{\infty} L(x_{n} + d + \gamma|\mathbf{x}_{(n-k+1,n)} )\;d\gamma.\quad\label{int1} 
\end{eqnarray}
\end{widetext}

Hence the joint PDF, the aposterior PDF and the total probability to find
increments are given by
\begin{widetext}
\begin{eqnarray}
j\bigl(\mathbf{x}_{(n-k+1,n)},Y_{n+1}(d)=1\bigr) & = & j\bigl(\mathbf{x}_{(n-k+1,n)}\bigr) \cdot L\bigl(Y_{n+1}(d)=1|\mathbf{x}_{(n-k+1,n)}\bigr)\\
\rho\bigl(\mathbf{x}_{(n-k+1,n)}|Y_{n+1}(d)=1\bigr) & = & \frac{j\bigl(\mathbf{x}_{(n-k+1,n)},Y_{n+1}(d)=1\bigr)}{P(Y(d)=1)},\\
P(Y(d)=1) &=& \int_{-\infty}^{\infty} dx_{n-k+1} \int_{-\infty}^{\infty} dx_{n-k+2} ... \int_{-\infty}^{\infty} dx_n \; j(\mathbf{x}_{(0,n-k)}\mathbf{x}_{(n-k+1,n)},Y_{n+1}(d)=1). \label{appendixptotal}
\end{eqnarray}
\end{widetext}

Whether we can acess a given stochastical process analytically or not depends on the question of whether the integrals in Eq.\ (\ref{appendixptotal}) can be solved or not.

If we are interested in the prediction of threshold crossings instead of
increments, we can interpret $\eta$ as the magnitude of the threshold and set
$x_n=0$ in order to obtain the corresponding expressions for the likelihood,
the joint PDF, the aposterior PDF, and the total probability.
\end{appendix}

%
\end{document}